\documentclass[]{spie}  

\usepackage{amsmath,amsfonts,amssymb}
\usepackage{graphicx}
\usepackage[colorlinks=true, allcolors=blue]{hyperref}

\title{Estimating Atmospheric Wind Speeds From Gemini Planet Imager AO Telemetry}
\author{{Zhenxi Du$^a$, Saavidra Perera$^a$, Daniel Levinstein$^a$, Quinn Konopacky$^a$, Alex Madurowicz$^b$, Bruce Macintosh$^b$, Lisa Poyneer$^c$, Richard Wilson$^d$, Ollie Farley$^d$}\\{\_}\\{$^a$University of California San Diego, 9500 Gilman Drive, La Jolla, CA 92093}\\{$^b$University of California Santa Cruz, 1156 High Street, Santa Cruz, CA 95064}\\{$^c$Lawrence Livermore National Lab, 7000 East Ave, Livermore, CA 94550}\\{$^d$Center for Advanced Instrumentation, University of Durham, South Rd, Durham DH1 3LS, United Kingdom}}
\date{June 2024}

\begin{document}
\maketitle
\begin{abstract}
The Earth's atmosphere is comprised of turbulent layers that result in speckled and blurry images from ground-based visible and infrared observations. Adaptive Optics (AO) systems are employed to measure the perturbed wavefront with a wavefront sensor (WFS) and correct for these distortions with a deformable mirror. Therefore, understanding and characterising the atmosphere is crucial for the design and functionality of AO systems. One parameter for characterizing the atmosphere is the atmospheric coherence time, which is a function of the effective wind velocity of the atmosphere. This parameter dictates how fast the AO system needs to correct for the atmosphere. If not fast enough, phenomena such as the wind butterfly effect can occur, hindering high-contrast coronographic imaging. This effect is a result of fast, strong, high-altitude turbulent layers. This paper presents two methods for estimating the effective wind velocity, using pseudo-open loop WFS slopes. The first method uses a spatial-temporal covariance map and the second uses the power spectral density of the defocus term. We show both simulated results and preliminary results from the Gemini Planet Imager AO telemetry. \\
\\
\textbf{Keywords:} Adaptive Optics, Atmospheric Profiling, Coherence time, AO Telemetry, Wind Butterfly Effect, Gemini Planet Imager
\end{abstract}

\section{Introduction}

Atmospheric turbulence poses a challenge for ground-based VIS/IR observations resulting in speckled and blurry images. Therefore, a deep understanding of the atmosphere is imperative. The atmosphere is comprised of multiple turbulent layers with varying turbulent strength and wind velocity. Adaptive optics (AO) systems are employed to correct the effects of the atmosphere. One way of characterising the atmosphere is by measuring the coherence time, which can be defined as 
\begin{equation}
   \tau_{0}=0.314\frac{r_{0}}{v_{\mathrm{eff}}}, 
\end{equation}
where $r_0$ is the Fried parameter and $v_{\mathrm{eff}}$ is defined as
\begin{equation} \label{2}
    v_{\mathrm{eff}}=\left[\frac{\int^{\infty}_{0}C^{2}_{n}(h)V(h)^{\frac{5}{3}}dh}{\int^{\infty}_{0}C^{2}_{n}(h)dh}\right]^{\frac{3}{5}},
\end{equation}
where $C^{2}_{n}(h)$ represents the strength of turbulence as a function of altitude, $h$, and $V(h)$ represents the velocity as a function of $h$ \cite{hardy}. This equation can be simplified to 
\begin{equation} \label{3}
    v_{\mathrm{eff}}^{\frac{5}{3}}=a_{\mathrm{0}}v_{\mathrm{0}}^{\frac{5}{3}}+...+a_{\mathrm{n}}v_{\mathrm{n}}^{\frac{5}{3}},
\end{equation}
where $a_{\mathrm{n}}$ denotes the relative strength of layer $n$ and $v_{\mathrm{n}}$ denotes the wind speed of that layer. Coherence time dictates the rate at which an AO system must correct the atmosphere. In high contrast imaging, if the rate of correction is too slow, the ``wind butterfly effect"\cite{2019JATIS...5d9003M} (also known as the ``wind-driven halo"\cite{refId0}) can be observed in coronagraphic images. This negatively impacts the achievable contrasts. This effect can be seen in a subsect of the Gemini Planet Imager (GPI) data. This effect occurs when fast strong high-altitude turbulence is present.

\indent Knowing the coherence time and wind profile is important for designing AO systems, understanding noise sources, predictive control and post-processing (e.g. removing the effects of the wind butterfly effect to improve contrasts). Work using GPI AO telemetry to measure the wind velocity of turbulent layers and understand frozen flow using Fourier modes has previously been shown\cite{Poyneer:09}. In this paper, we will discuss two different methods of analyzing AO telemetry data to estimate $v_{\mathrm{eff}}$. The Fried parameter, which is also required to calculate the atmospheric coherence time, is not covered in this paper, however, it can be estimated using similar methods employed by profiling instruments SLODAR \cite{10.1046/j.1365-8711.2002.05847.x} or SHIMM \cite{10.1093/mnras/stad339}. In section \ref{sec:methods}, we discuss the methods for estimating $v_\mathrm{eff}$ using simulated POL data from the soapy Python library\cite{10.1117/12.2232438}. In section \ref{sec:results} we present simulated results, as well as preliminary results from GPI AO data.


\section{Methods}
\label{sec:methods}
As described by Equation \ref{2}, the relative strengths and wind velocities of each turbulent layer are needed for estimating the effective wind velocity. We present two different methods for estimating this parameter: (i) spatial-temporal covariance maps and (ii) power spectral distribution (PSD) of the defocus Zernike term. We simulated POL slopes based on the GPI AO system (see Table \ref{tab:1}) using the soapy Python library. 

\begin{table}[h]
    \centering
    \begin{tabular}{|c|c|}\hline
 \textbf{Parameter} & \textbf{Value}\\\hline   
        WFS &Shack Hartmann\\\hline
        Pupil diameter*& 7.77 m\\ \hline 
        Central obscuration$^\dagger$& 1.29 m\\ \hline
        Frame rate&1 kHz\\\hline
 Subaperture array&43 x 43\\\hline
    \end{tabular}
    \caption{AO system configuration. *Set by the undersized secondary mirror and a further-undersized mask on the secondary. $^\dagger$Set by the hole in the center of the secondary mirror.\cite{Macintosh_2014} }
    \label{tab:1}
\end{table}

\subsection{Method 1: Spatial temporal covariance map}
\begin{figure}
    \centering
    \includegraphics[width=0.3\linewidth]{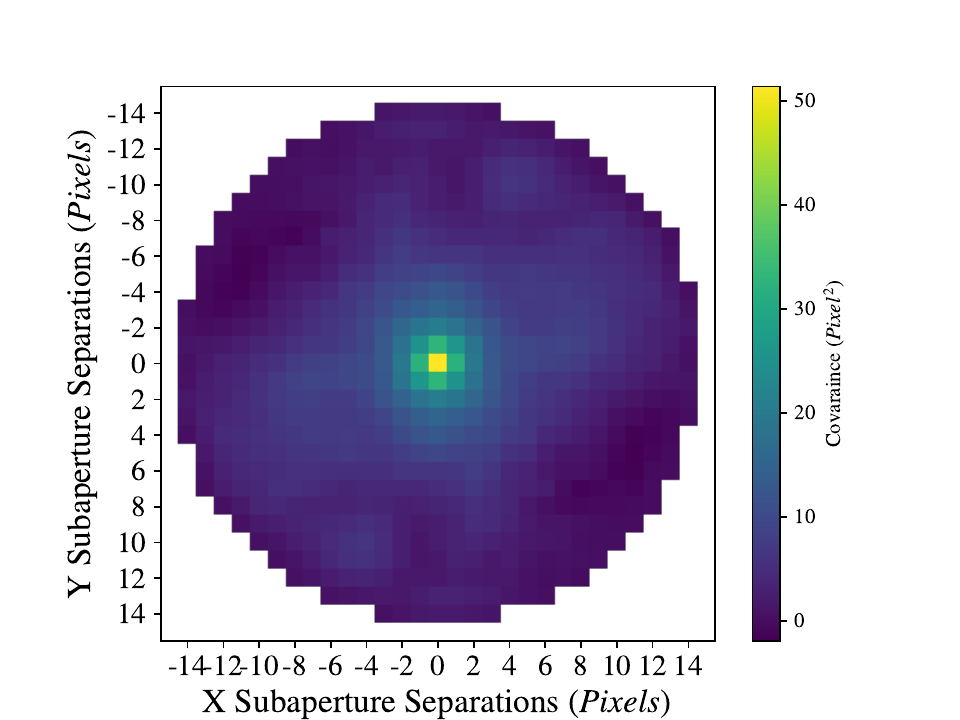}
    \includegraphics[width=0.3\linewidth]{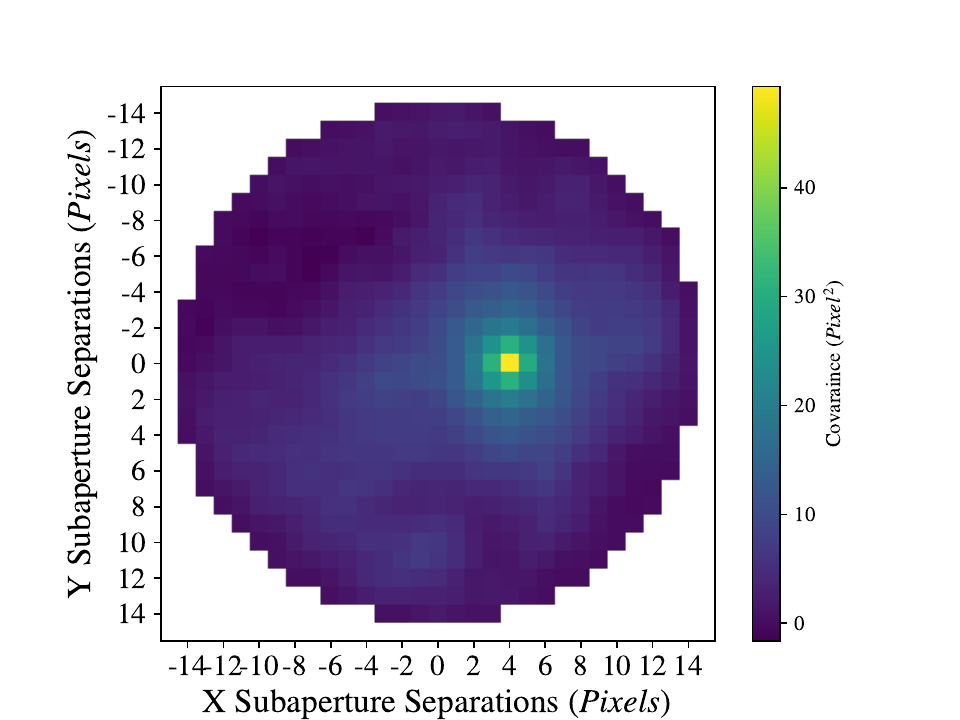}
    \includegraphics[width=0.3\linewidth]{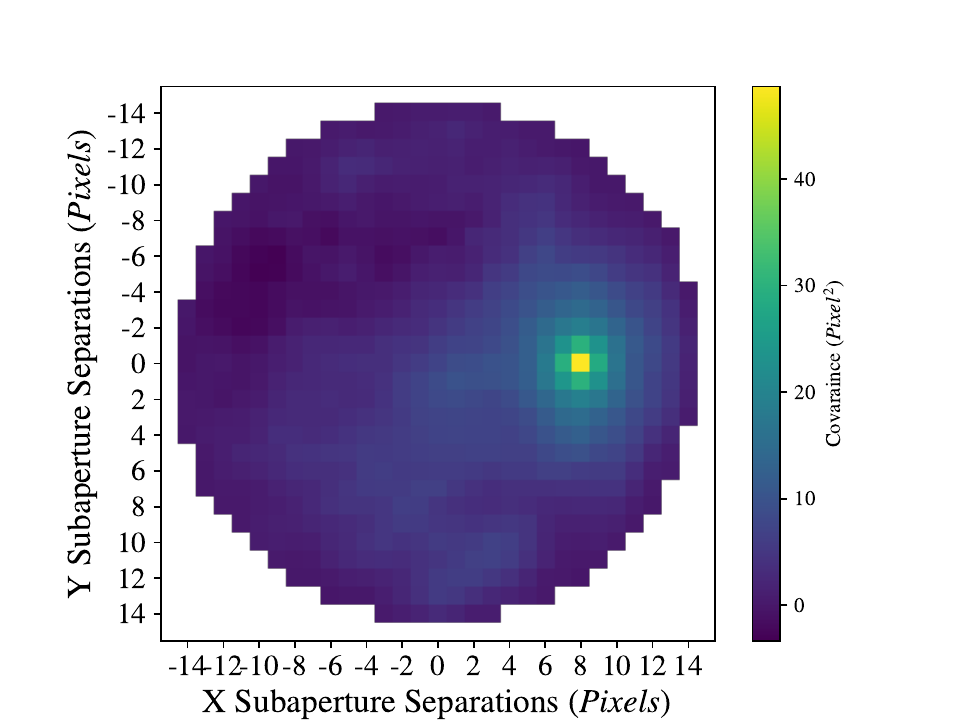}
    \caption{Covariance map for a single layer turbulence profile, with a wind speed of 20 m/s. [Left] Covariance map with no temporal offset. [Middle] Covariance map with 100 frames (0.1s) temporal offset. [Right] Covariance map with 200 frames (0.2s) temporal offset.}
    \label{fig:one_layer_cov}
\end{figure}

One way to identify the strengths and velocities of turbulent layers is by constructing a spatial-temporal covariance map showing the covariance of POL WFS slopes as a function of subaperature separation for a given temporal offset. This is calculated by
\begin{equation} A_{\delta{i},\delta{j},\delta{t}}=\langle C_{i,j,t}C_{i^{'},j^{'},t^{'}}^{'} \rangle,
\end{equation}
where $C$ and $C^{'}$ are WFS slopes of positions $[i,j]$ and $[i^{'},j^{'}]$ at times $t$ and $t^{'}$, $[\delta{i},\delta{j}]$ represent the subaperture separation and $\delta{t}$ represent the temporal offset. At $\delta{t}$ = 0, all layers are superimposed in the centre of the map. By increasing $\delta{t}$ all turbulent layers will traverse the covariance map as peaks \cite{2009JOSAA..26..833P} according to their velocity, as seen in Figure \ref{fig:one_layer_cov}. The computation of the covariance map for 43x43 slopes can take a long time to calculate, therefore they are binned to reduce the dimension down to 14x14. This method and reduction is presented by Levinstein et al. (2022)\cite{Levinstein_2022}. 

The locations of these peaks will indicate the direction and speed of these turbulent layers by knowing the temporal offset and subaperture sizes\cite{perera}\cite{10.1093/mnras/stab3813}. For example, in Figure \ref{fig:one_layer_cov} (middle) the peak is displaced by 4 subapertures over 100 frames (0.1 s), therefore the speed is 22.2 m/s (the input wind speed is 20 m/s).

\begin{figure}
    \centering
    \includegraphics[width=0.45\linewidth]{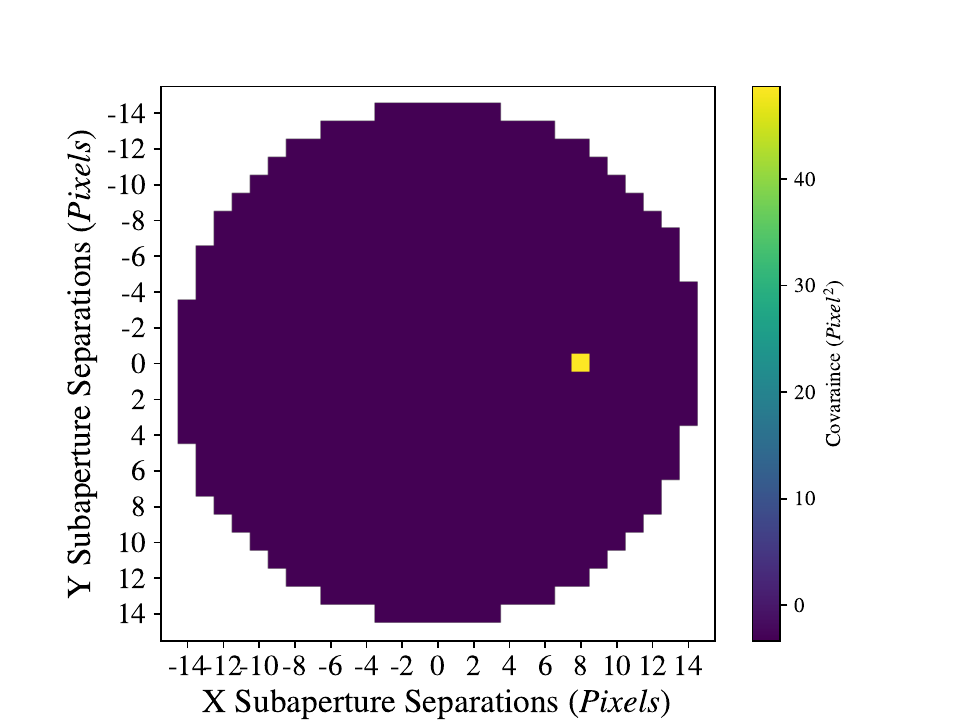}
    \includegraphics[width=0.45\linewidth]{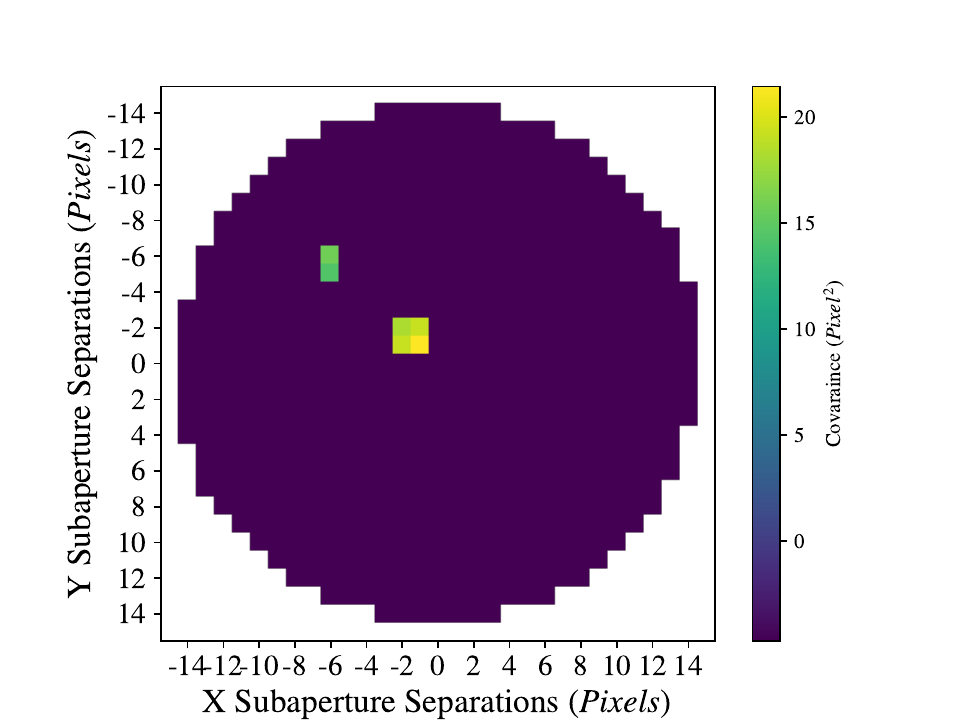}
    \caption{Covariance map with thresholding of 0.15 and a temporal offset of 200 frames (0.2s). [Left] Simulated WFS data for a single layer turbulence, with a wind speed of 20 m/s. [Right] Simulated WFS data for a two-layer turbulence profile, with wind speeds of 5 and 20 m/s and relative strengths of 0.5 and 0.5 respectively.}
    \label{fig:thresh}
\end{figure}

\indent To identify the turbulent layers, thresholding is first applied to the covariance map, whereby anything below this threshold is ignored. For the simulated data, the thresholding constant was set to 0.1 $\times$ the maximum value of the map. After thresholding, each turbulent layers are separated into distinct regions (an example can be seen in Figure \ref{fig:thresh}), and the exact locations of these layers are calculated from the relative strength and position of each pixel in these regions. For a given $\delta{t}$ the covariance map can essentially be turned into a velocity map to calculate the effective wind speed. Since the temporal offset and subaperture size are known, every point on the covariance map can be converted to a velocity based on how many subaperture separations it is. This velocity map is then matched with the covariance map, where the covariance at each point can be translated to the relative strengths. With the wind speeds and relative strengths, the effective wind speed can be calculated with Equation \ref{3}.

\subsection{Method 2: Defocus Power spectral density}
\begin{figure}
    \centering
    \includegraphics[width=0.45\linewidth]{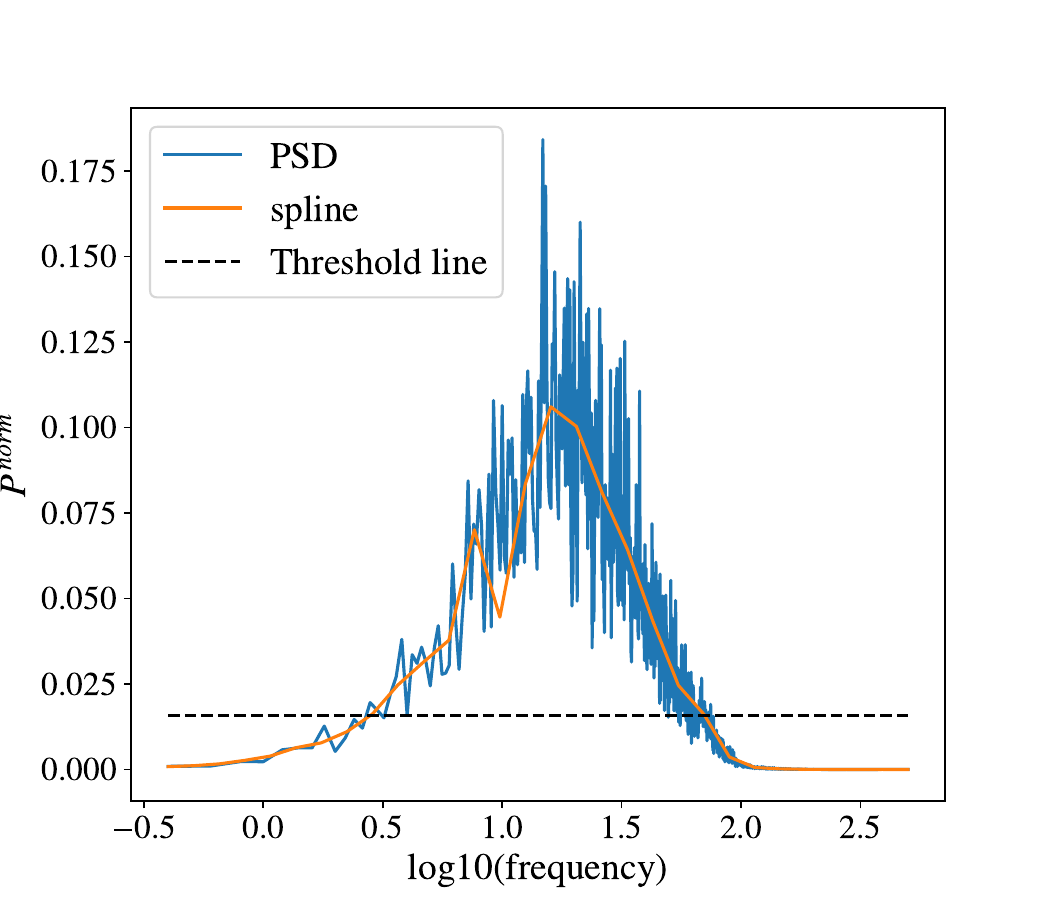}
    \caption{PSD constructed with the same set of one layer WFS data as Figure \ref{fig:one_layer_cov}.}
    \label{fig:one_layer_pow}
\end{figure}

Another method of estimating the effective wind velocity is to look at the PSD of the slopes. We can identify the wind velocity by locating the knee frequency of the PSD \cite{Hogge}. However, the knee frequency is hard to estimate in noisy, multi-layered data. Thus we look at the defocus Zernike mode. Two other modes (Tip/Tilt) that are stronger than defocus exist. However, they are not spherically symmetric and will therefore change with wind direction. The concept of using the defocus term to measure coherence time has been used before with DEfocus \cite{tokovinin} and the SHIMM\cite{10.1093/mnras/stad339}. The method presented here is from the SHIMM. The PSD is normalized such that 
\begin{equation}
    P^{norm}=\frac{f\Phi(f)}{\int\Phi(f)df},
\end{equation}
where $\Phi$ is the PSD of the defocus term and $f$ is frequency. Figure \ref{fig:one_layer_pow} shows for a single turbulent layer the PSD will peak at a frequency which is related to wind speed as 
\begin{equation}
    f_{peak} = \gamma\frac{v}{d},
\end{equation}
where $\gamma$ is an empirically determined constant and $d$ is the subaperture size.  By normalizing the PSD, the area under the curve is equal to one. Therefore, each frequency (i.e. velocity) has a relative turbulent strength associated with it. By using Equation \ref{3} $v_\mathrm{eff}$ can be calculated.   

\begin{figure}[bh]
    \centering
    \includegraphics[width=0.45\textwidth]{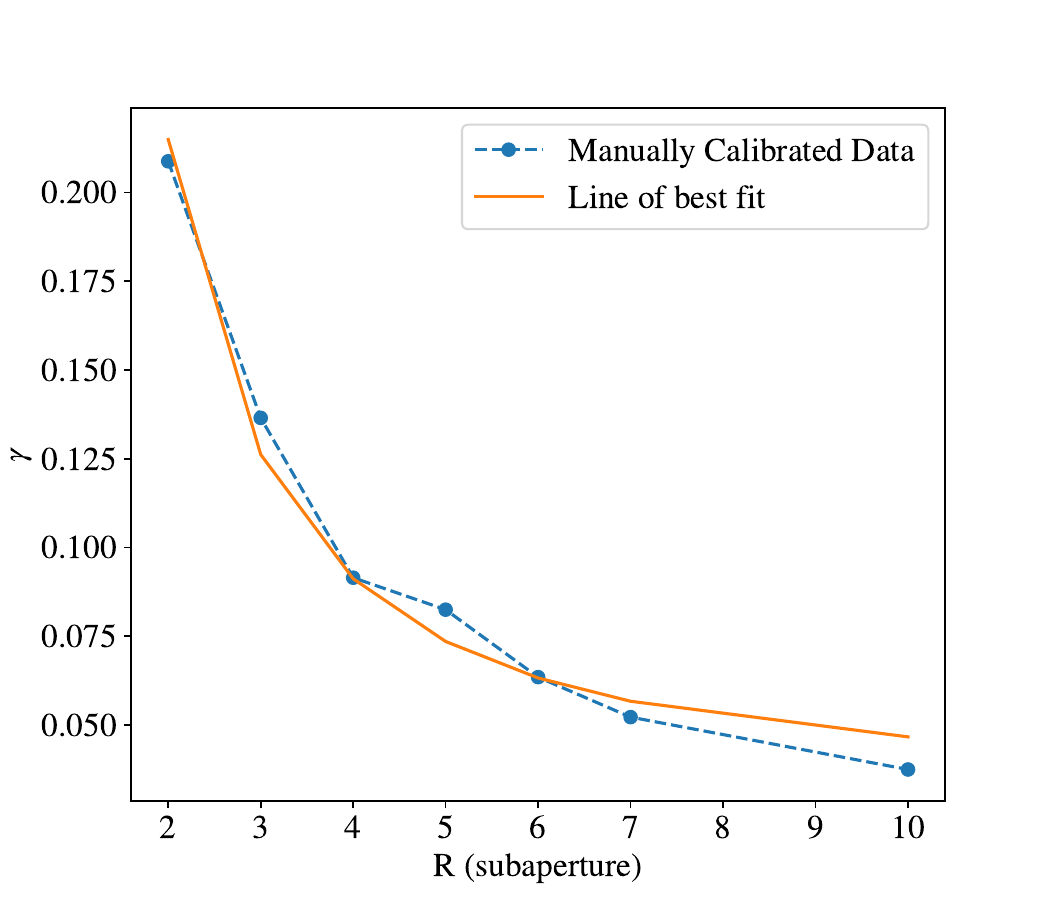}
    \caption{Plotted relation between $\gamma$ and sampling radius, with simulated data (blue) and fit (orange).}
    \label{fig:r_gamma}
\end{figure}

For this method, the size of the telescope needs to be considered. If the telescope is too large, all layers move to lower frequencies making it harder to distinguish between wind speeds,  but if it is too small faster layers will move to higher frequencies that can not be measured at 1 kHz frame rate. Therefore, the sample area is artificially restricted, and the ideal sample area was determined from simulated WFS data. We determined that a sampling diameter of 0.77 m at a frame rate of 1 kHz, yields the most accurate effective wind velocity. We also found that the constant $\gamma$ is related to the radius of the telescope, $R$,
\begin{equation}
    \gamma \propto R^{-\frac{5}{3}}.
\end{equation}
\noindent This relationship is shown in Figure \ref{fig:r_gamma}. To reduce the noise of the power spectrum, a moving average was first applied to the PSD, followed by a spline interpolation to further smooth the data, this is depicted by the orange line in Figure \ref{fig:one_layer_pow}. Lastly, a threshold of 0.15 $\times$ the maximum value is applied to reduce noise.

\section{Results}
\label{sec:results}
\subsection{Simulated Results}
\begin{figure}
    \centering
    \includegraphics[width=0.3\linewidth]{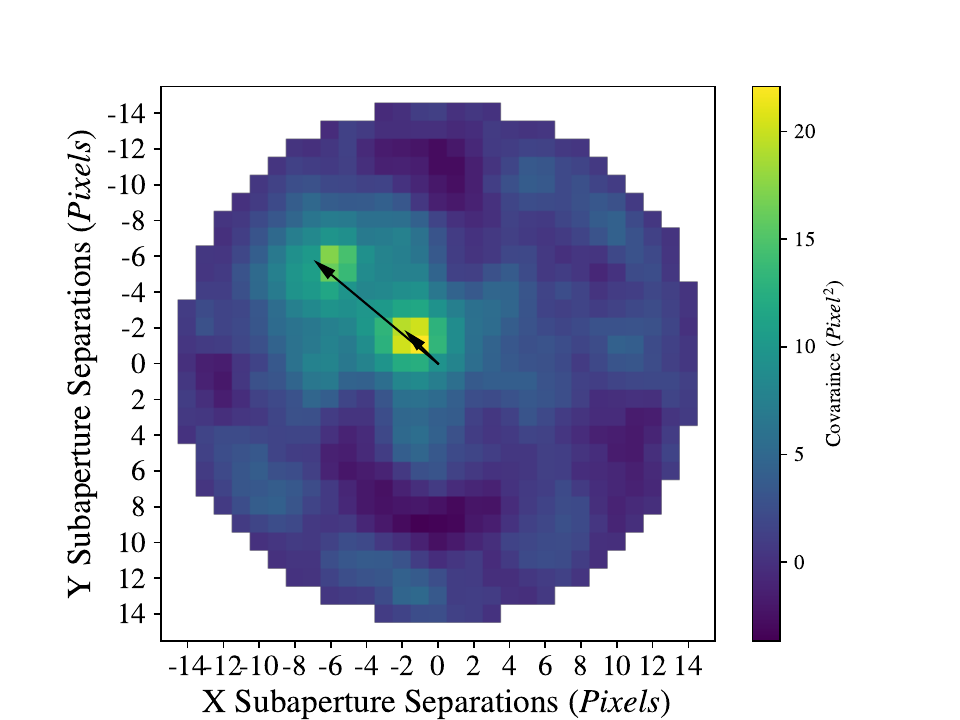}
    \includegraphics[width=0.3\linewidth]{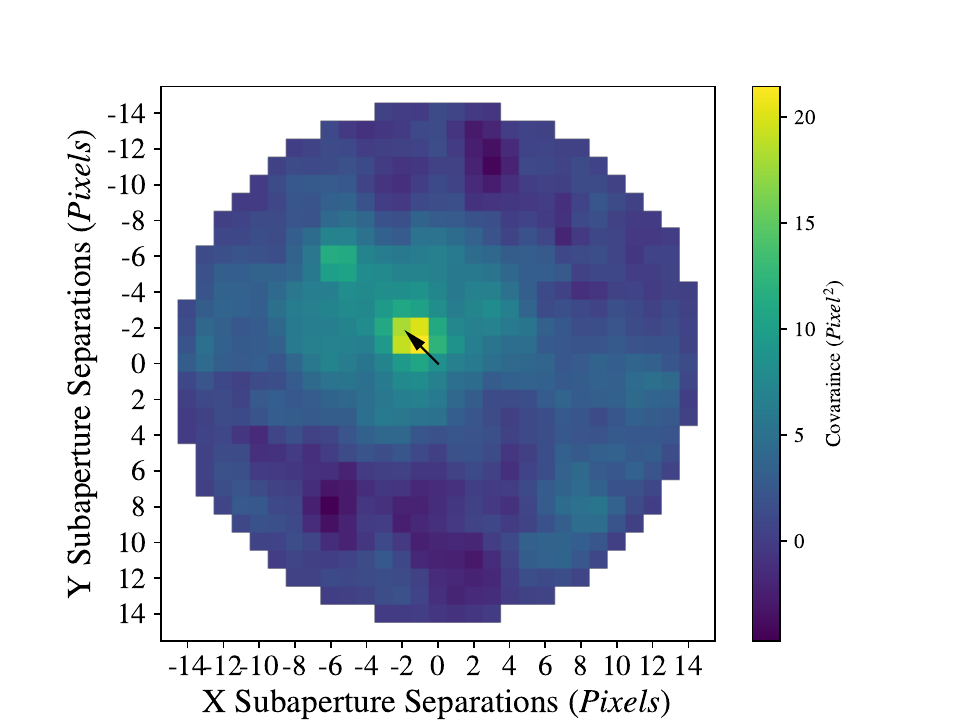}
    \includegraphics[width=0.3\linewidth]{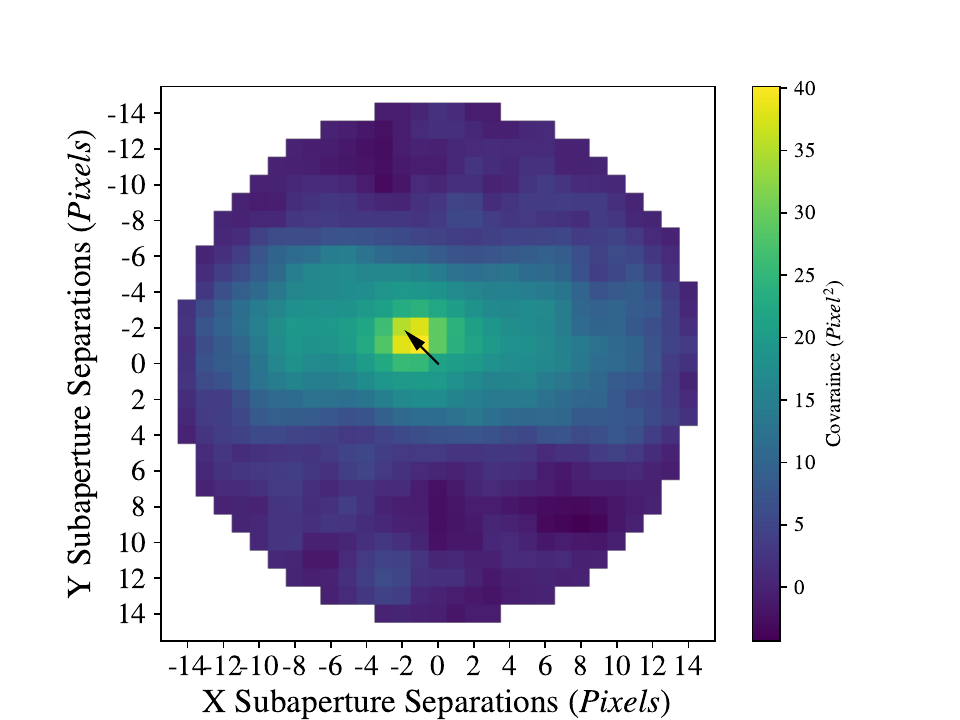}
    \includegraphics[width=0.3\linewidth]{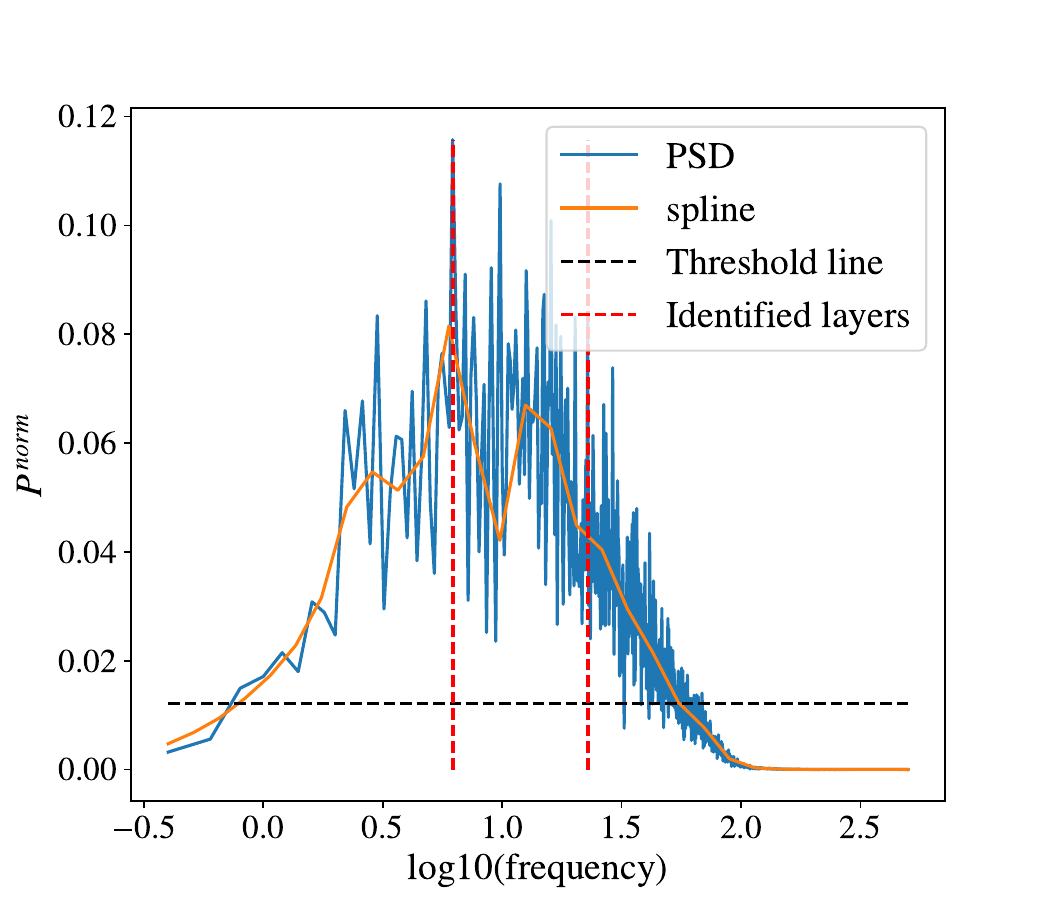}
    \includegraphics[width=0.3\linewidth]{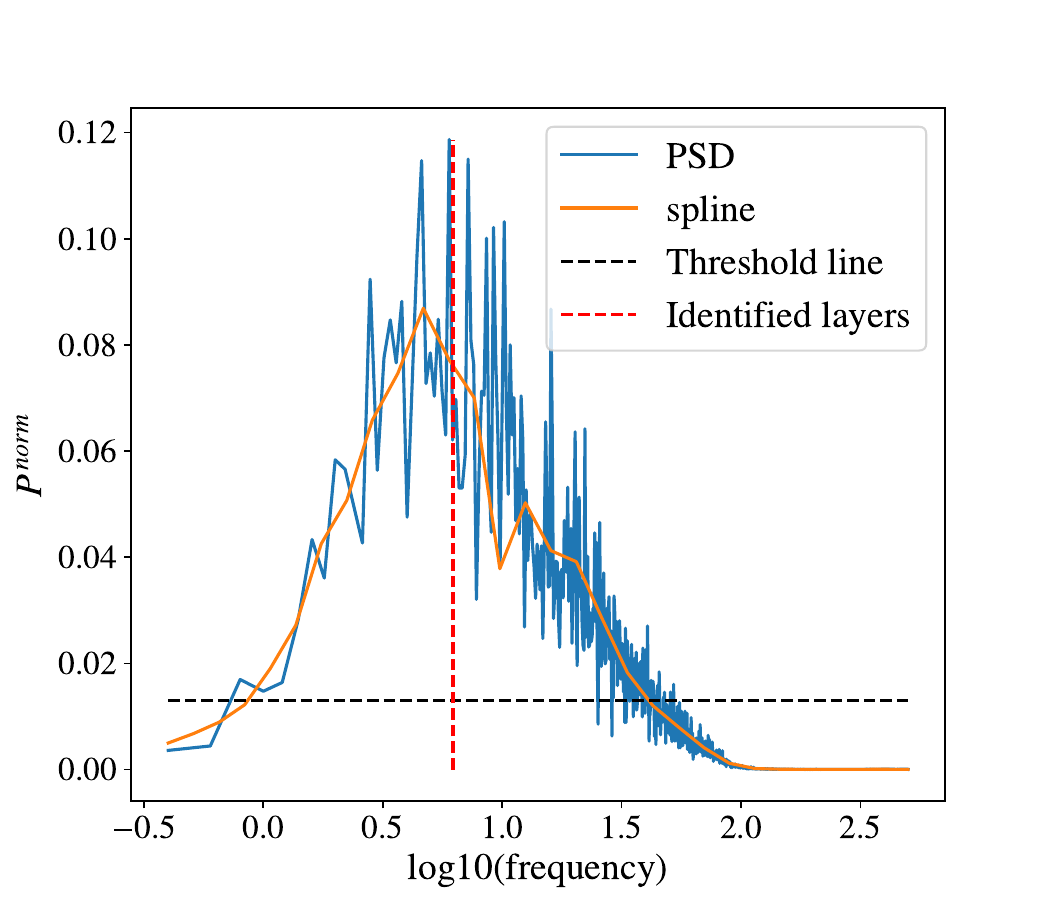}
    \includegraphics[width=0.3\linewidth]{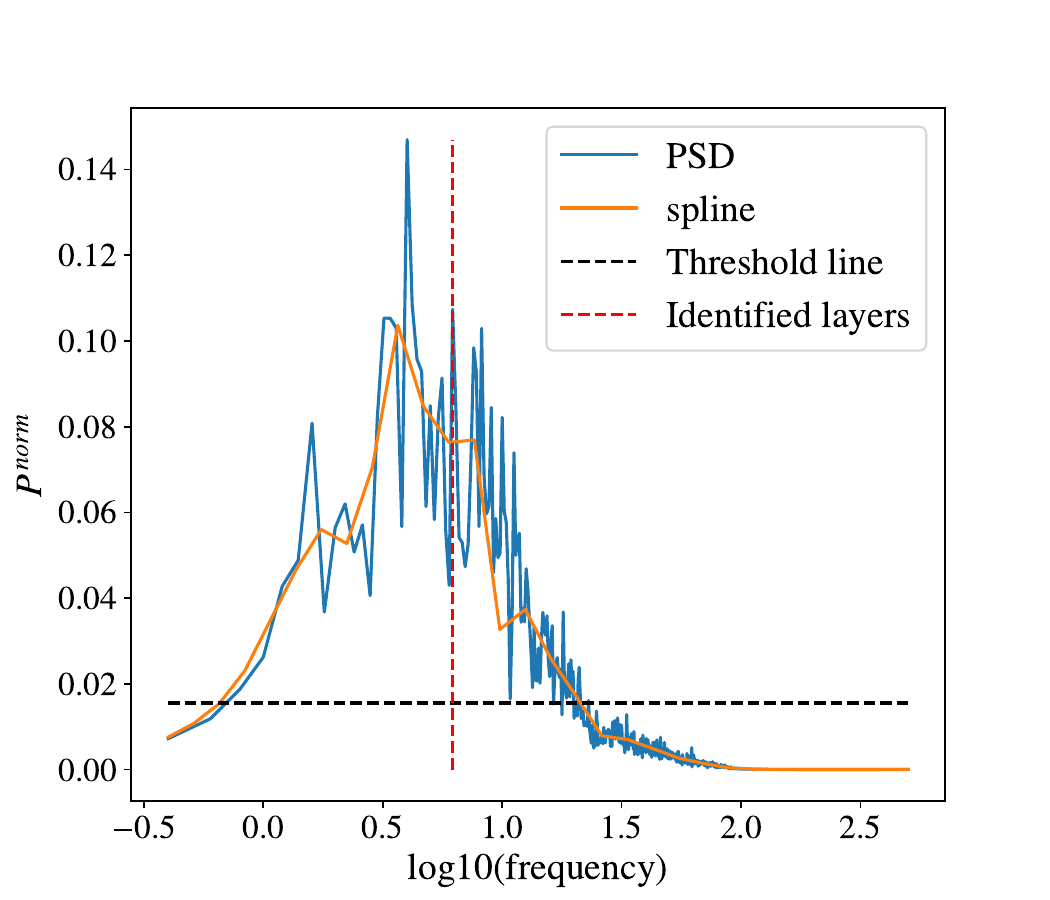}
    \caption{[Top] Covariance map retrieved from a simulated two-layer profile, with wind speeds of 5 and 20 m/s of different relative strengths. Layers are identified in the covariance maps with arrows. [Bottom] Power spectrum with data reduction methods applied. Red dashed lines indicate the locations of turbulent layers identified by the covariance map. Each set of covariance map/PSD pairs is constructed with the same simulated WFS data. [Left] Relative strength of 0.5/0.5. Expected $v_{eff}$ = 13.97 m/s. Method 1 and 2 give $\sim$10.90 m/s and $\sim$12.55 m/s respectively. [Middle] Relative strength of 0.7/0.3. Expected $v_{eff}$ = 11.00 m/s. Method 1 and 2 give $\sim$5.31 m/s and $\sim$8.72 m/s respectively. [Right] Relative strength of 0.9/0.1. Expected $v_{eff}$ = 7.37 m/s. Method 1 and 2 give $\sim$5.30 m/s and $\sim$5.46 m/s respectively. }
    \label{fig:two_layer}
\end{figure}

\begin{figure}[!tbp]
  \centering
    \includegraphics[width=0.5\textwidth]{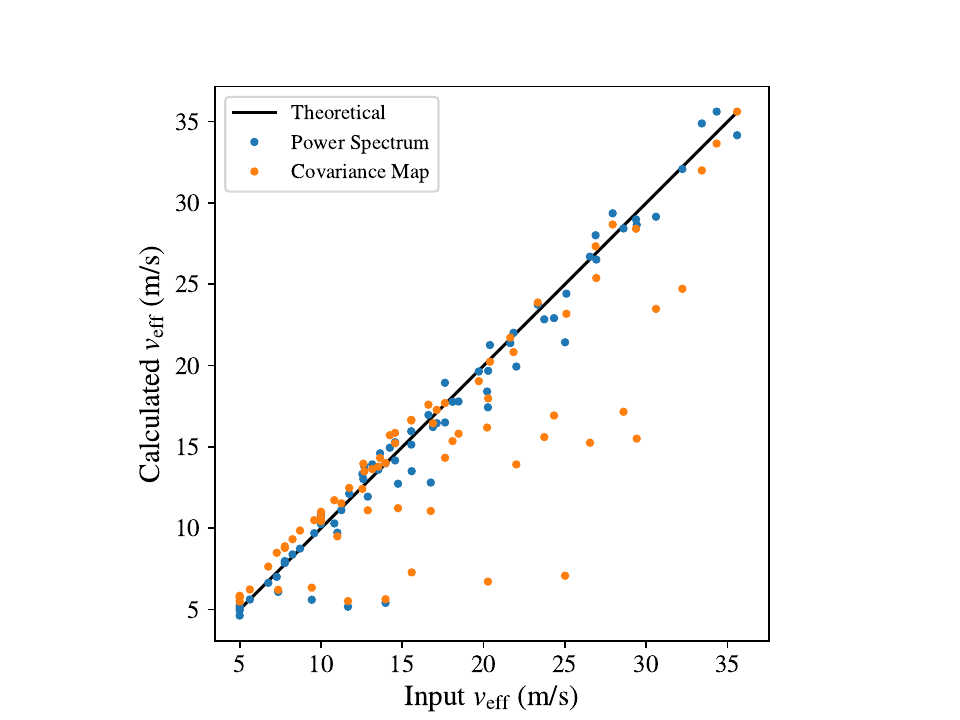}
    \caption{Comparison of calculated effective wind velocity and input simulated wind velocity for power spectrum and covariance map.}
    \label{fig:scatter}
\end{figure}

Both methods were tested in simulation for a two-layer turbulent atmosphere, (i) layer one with wind speed of 5 - 20 m/s and relative turbulent strength of 0.5, 0.6, 0.7 and 0.9 (ii) layer two with wind speeds between  5 - 50 m/s and relative turbulent strength of 0.5, 0.4, 0.3 and 0.1. Figure \ref{fig:two_layer} shows an example covariance map and power spectrum from simulated WFS data.
Figure \ref{fig:scatter} shows results comparing the input and calculated $v_\mathrm{eff}$. Both methods show promising results with the PSD method showing a stronger correlation than the covariance map method. The majority of the outlier results that deviate from the theoretical result are data sets containing a fast but weak wind layer. This is because the layers become below the thresholding value used to remove noise. As such, the effective wind velocities estimated by both methods are expected to be lower than the actual value in these cases. However, this effect is much weaker in the PSD method. For example, in Figure \ref{fig:two_layer} (middle), a peak is barely visible around the location where the second (20 m/s) layer is expected to be. However, this peak is not identified because it is flagged as noise and removed in the thresholding. 
The PSD method is also computationally faster than the covariance mapping algorithm. However, unlike the covariance map method, it is unable to distinguish the direction of wind layers and individual turbulent layers.\\

\subsection{Comparison of Methods on GPI Data}
\label{sec:GPI}
\begin{figure}
    \centering
    \includegraphics[width=0.3\linewidth]{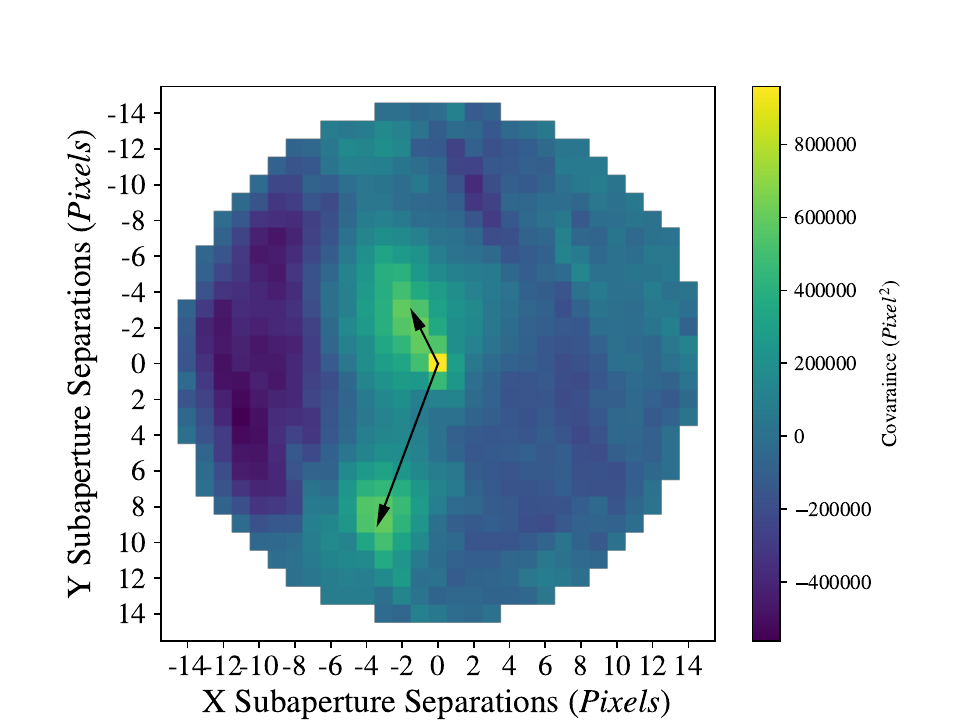}
    \includegraphics[width=0.3\linewidth]{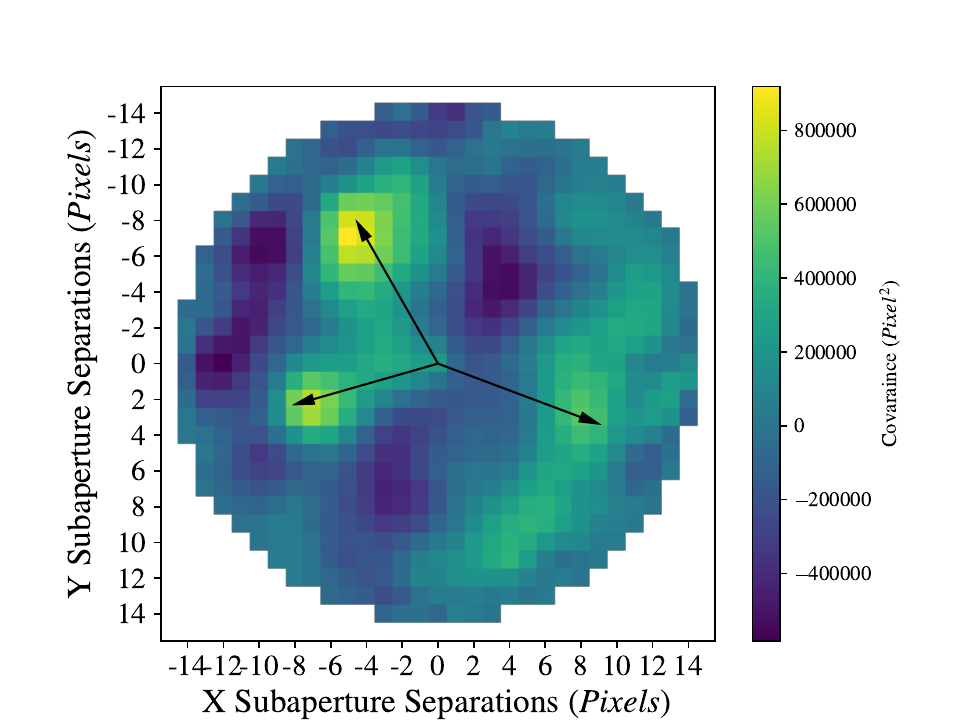}
    \includegraphics[width=0.3\linewidth]{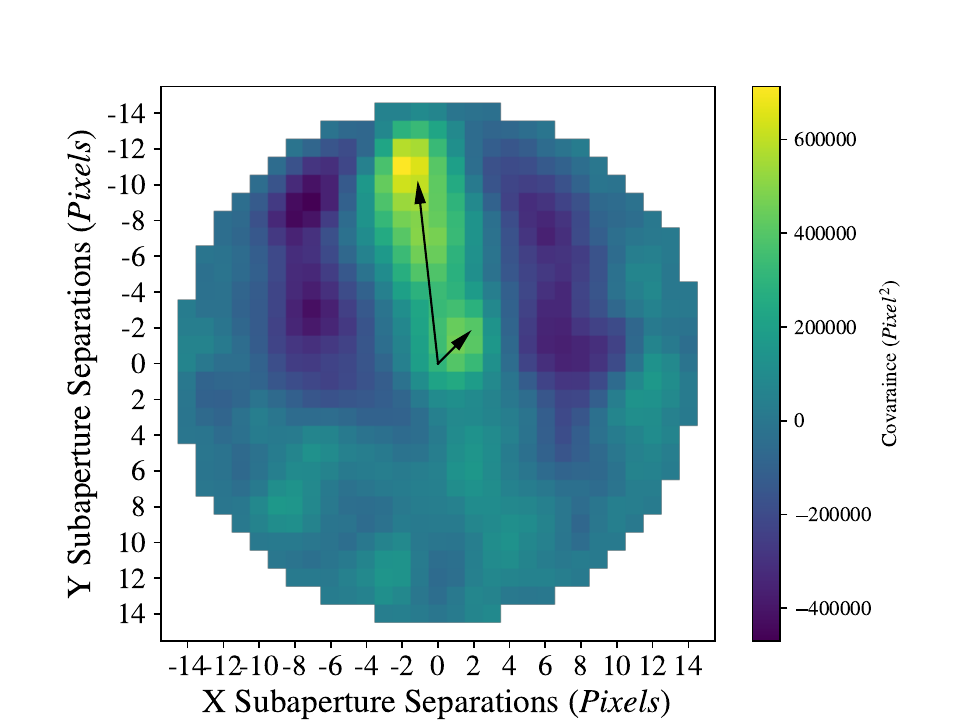}
    \includegraphics[width=0.3\linewidth]{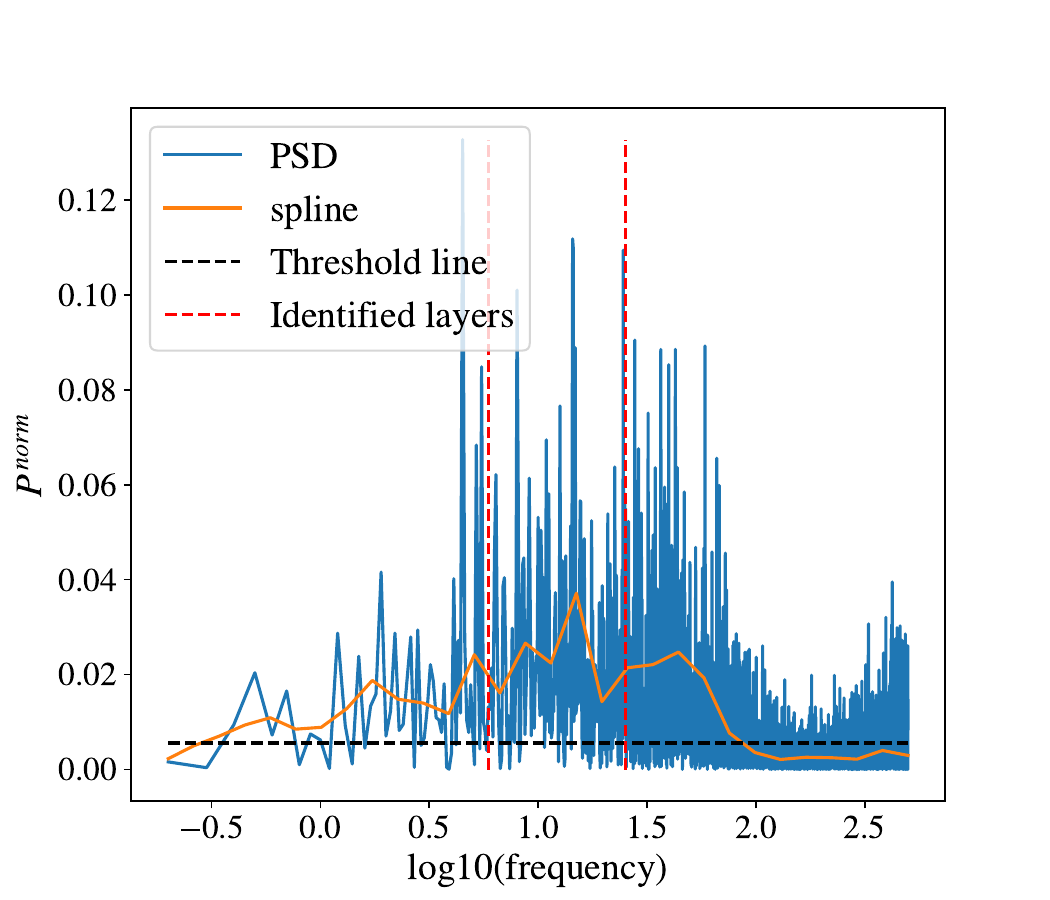}
    \includegraphics[width=0.3\linewidth]{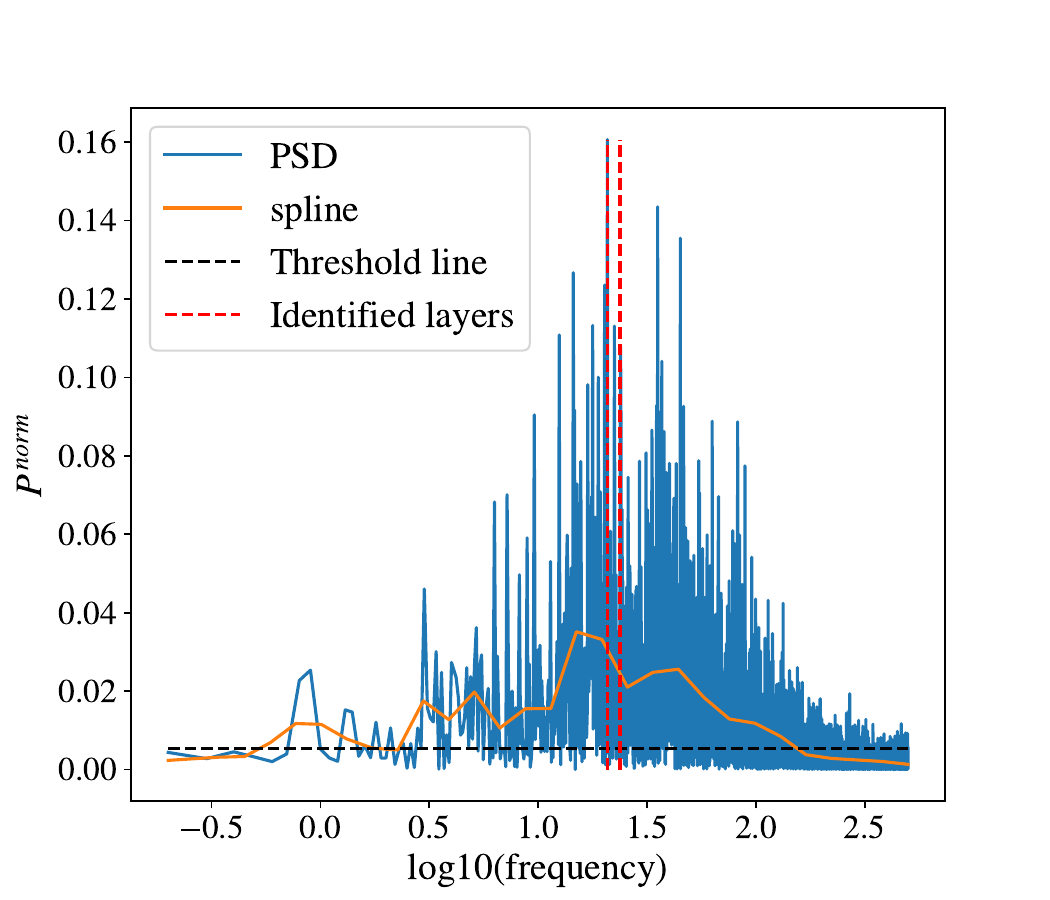}
    \includegraphics[width=0.3\linewidth]{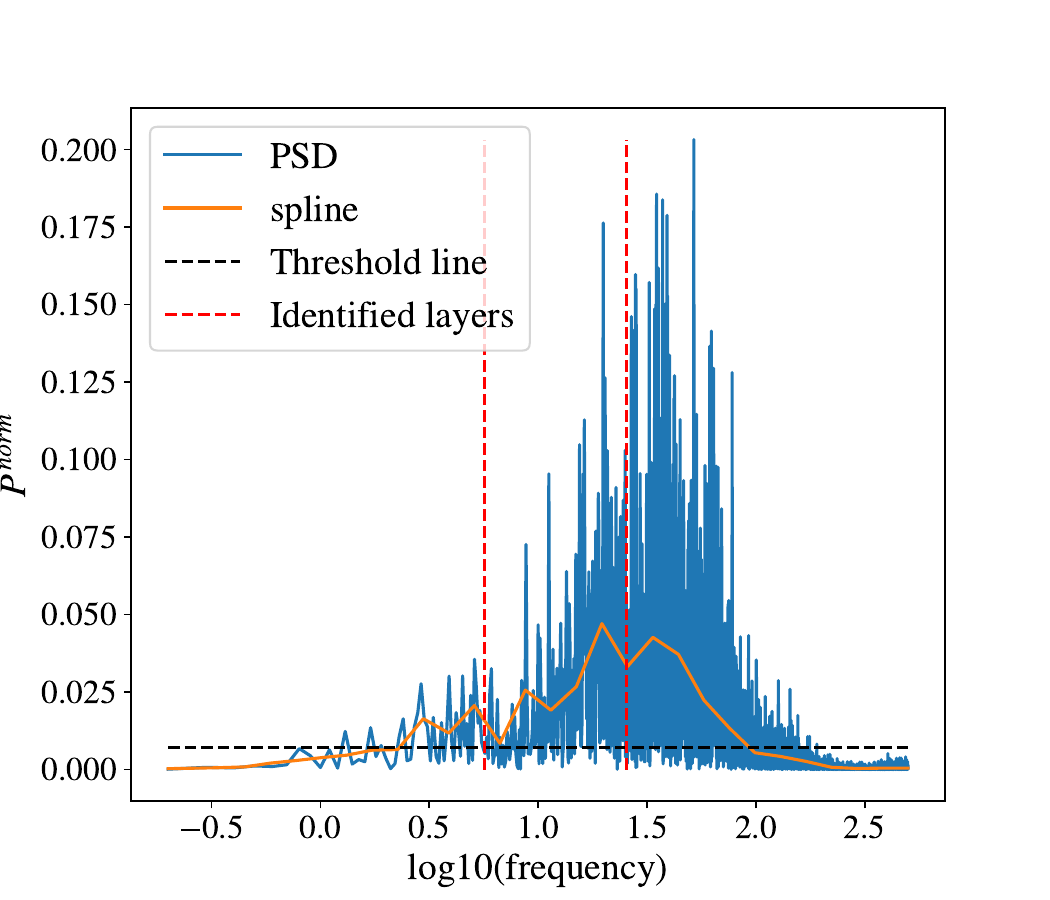}
    \caption{[Top] Covariance map with a temporal offset of 0.2s. Distinct turbulence layers are located and plotted out with arrows. [Bottom] Power spectrum with data reduction methods applied. Red dashed lines indicate the locations of turbulent layers identified by the covariance map. Each set of covariance map/PSD pairs are constructed with the same WFS data from GPI. [Left] Method 1: $\sim$11.90 m/s $v_{eff}$, method 2: $\sim$19.24 m/s $v_{eff}$. [Middle] Method 1: $\sim$19.75 m/s $v_{eff}$, method 2: $\sim$26.22 m/s $v_{eff}$ (Note: Two layers of very similar velocities were identified by the covariance map and since the PSD disregard direction, the two layers' dashed lines were plotted on top of one another). [Right] Method 1: $\sim$20.46 m/s $v_{eff}$, method 2: $\sim$25.66 m/s $v_{eff}$.}
    \label{fig:real}
\end{figure}
\indent After calibrating the two methods on simulated WFS data, they were applied to GPI WFS data. Examples can be seen in Figure \ref{fig:real}. The effective wind velocity results corroborate with the simulated results, where the PSD yields a higher velocity due to it being more sensitive to weaker, faster layers. 
It is worth noting a spike in the PSD can be observed in very high frequency in real data (e.g. left panel of Figure \ref{fig:real}), the algorithm ignores this portion of the PSD as its velocity is too high to be a turbulent layer (v $>$ 150 m/s) and it is most likely an artefact of photon noise.

\section{Conclusion}
Both of our methods have shown promising results on both simulated and GPI WFS data. Overall, the simulated result suggests the PSD method is better at estimating the effective wind speed, with more accuracy and faster processing speed. However, the covariance map method provides more information about the turbulent layers and proves to be better at identifying individual strong layers. The next step is to apply the algorithms to more sets of GPI data to collect more data on the performance of these methods. By collecting effective wind speed measurements over all GPI AO telemetry datasets we can use this to build site statistics for coherence time and additionally compare this to coronagraphic data to further characterize the wind butterfly effect and possibly remove the effect post-processing. Improvements can be made to both algorithms to provide faster processing.

\begin{acknowledgements}
The GPI project has been supported by Gemini Observatory, which is operated by AURA, Inc., under a cooperative agreement with the NSF on behalf of the Gemini partnership: the NSF (USA), the National Research Council (Canada), CONICYT (Chile), the Australian Research Council (Australia), MCTI (Brazil) and MIN CYT (Argentina).
\end{acknowledgements}
\bibliography{report}
\bibliographystyle{spiebib}

\end{document}